\documentclass[%
reprint,
superscriptaddress,
showpacs,
amsmath,amssymb,amsfonts,
aps,
prb,
floatfix,notitlepage,latexsym
]{revtex4-1}

\usepackage[dvipdfmx]{graphicx}
\usepackage{dcolumn}
\usepackage{bm}
\usepackage{braket}
\usepackage{mediabb} 
\usepackage[dvipdfmx]{attachfile2}

\begin{abstract}
Controlling electric and magnetic properties of matter by laser beams is actively explored in the broad region of condensed matter physics, including spintronics and magneto-optics. Here we theoretically propose an application of optical and electron vortex beams carrying intrinsic orbital angular momentum to chiral ferro- and antiferro- magnets. We analyze the time evolution of spins in chiral magnets under irradiation of vortex beams, by using the stochastic Landau-Lifshitz-Gilbert equation. We show that beam-driven nonuniform temperature lead to a class of ring-shaped magnetic defects, what we call skyrmion multiplex, as well as conventional skyrmions. We discuss the proper beam parameters and the optimal way of applying the beams for the creation of these topological defects. Our findings provide an ultrafast scheme of generating topological magnetic defects in a way applicable to both metallic and insulating chiral (anti-) ferromagnets.

\vspace{3mm}
\noindent PhySH:Angular momentum of light, Ultrafast phenomena, LLG, Skyrmions
\end{abstract}

\begin{document}


\title{Ultrafast generation of skyrmionic defects with vortex beams:\\ printing laser profiles on magnets}
\author{Hiroyuki Fujita}
\thanks{Corresponding author}
\affiliation{Institute for Solid State Physics, University of Tokyo, Kashiwa 277-8581, Japan}
\affiliation{Kavli Institute for Theoretical Physics, University of California, Santa Barbara, California 93106, USA}
\email{h-fujita@issp.u-tokyo.ac.jp}
\author{Masahiro Sato}
\affiliation{Department of Physics, Ibaraki University, 
Mito, Ibaraki 310-8512, Japan}
\affiliation{Spin Quantum Rectification Project, ERATO, Japan Science and Technology Agency, Sendai
980-8577, Japan}
\email{masahiro.sato.phys@vc.ibaraki.ac.jp}
\date{\today}

\maketitle
\onecolumngrid
\section{Introduction}
The optical vortex, first proposed in 1992~\cite{PhysRevA.45.8185,PhysRevLett.88.053601}, is an electromagnetic wave carrying intrinsic orbital angular momentum (OAM). Although often confused with circularly polarized light which has a finite spin angular momentum of photons, they are different concepts.
The intrinsic OAM twists the phase structure of the propagating beams and forces them to have topological singularity along the propagation axis, a line with vanishing beam intensity. As a result, optical vortices have a ring-shaped spatial profile of the intensity. More recently, beams of electrons with intrinsic OAM were also proposed and experimentally realized~\cite{PhysRevLett.99.190404,Uchida:2010aa,Verbeeck:2014aa} whose intensity is spatially ring-shaped as well. In this paper, we call those beams with OAM ``vortex beams".

Applications of optical vortices are actively explored. We can transfer their OAM to classical particles~\cite{PhysRevLett.75.826,Friese:1998aa,Paterson:2001aa} or excitons~\cite{PhysRevB.93.045205} to induce rotational motion of them or use their phase structure to realize super-resolution microscopy~\cite{PhysRevLett.98.218103} and chiral laser ablation~\cite{Hamazaki:10,Omatsu:10,Toyoda:2012aa,Takahashi:13,PhysRevLett.110.143603,Litchinitser1054} to name a few. 
%
%
%
Compared to those of optical vortices, applications of electron vortex beams have been not so much discussed yet, but, for example, we can use them to achieve direct imaging of electron Landau levels~\cite{Schattschneider:2014aa}.

Regarding those applications of vortex beams as ``printing" of the spatial profile (phase or intensity) of the beams to physical systems, as a natural extension of them, we hit on the possibility of their microscopic analogs, {\it i.e.} encoding the profile of vortex beams into solids by using electronic or magnetic degrees of freedoms. However, such applications for solid state physics are so far unexplored. In this paper, we particularly focus on spin systems and consider a use of vortex beams for magnetism. 

The study of the interaction between lights and magnets goes back to Faraday's era, but it is only very recently that one can exploit intense lasers in the broad frequency region and observe microscopic magnetic textures in ultrafast ways. Nowadays, such laser-based ultrafast magnetic dynamics is a hot topic in condensed matter physics~\cite{PhysRevLett.76.4250,Kimel:2005aa,Bigot:2009ab,PhysRevLett.105.147202,RevModPhys.82.2731,Kampfrath:2011aa,Ostler:2012aa,Satoh:2012aa,PhysRevB.88.220401,arXiv:1404.2010,Mangin:2014aa,PhysRevB.90.085150,Carva:2014aa,PhysRevB.90.214413,Mentink:2015aa,Mikhaylovskiy:2015aa,wilson2016,arXiv:1602.03702,:/content/aip/journal/apl/108/1/10.1063/1.4939447,1367-2630-18-1-013045,R.:2016aa}, especially in spintronics~\cite{spincurrentbook} and magneto-optics~\cite{magop_book,Kimel:2005aa,Ghamsari:2016aa}. Here, as a first example of such ``magneto-singular-optics", we work on chiral ferro- and antiferromagnets, which have strong Dzyaloshinskii-Moriya (DM) interaction~\cite{DZYALOSHINSKY1958241, PhysRev.120.91} and host topological magnetic defects called skyrmions~\cite{JETPLett.222451975,Yablonskii1989, Ivanov:1990aa, Hubert1994,Bogdanov2006,Leonov2016,Fert:2013aa}. 

Isolated skyrmions in chiral ferromagnets (FMs) are prospected as a promising candidate as bits for future magnetic memory devices with low energy consumption~\cite{Seki198,Yu:2012aa,Nagaosa:2013aa,Tomasello:2014aa}. More recently, skyrmions in antiferromagnets (AFMs) are also actively discussed~\cite{BaryakhtarJETP,Wolf2002,PhysRevLett.116.147203, Zhang:2016aa,1367-2630-18-7-075016} in the context of antiferromagnetic spintronics. There are many theoretical proposals on the creation of isolated skyrmions in chiral ferromagnets~\cite{PhysRevB.85.174416,SampaioJ.:2013aa,Zhou:2014aa,Koshibae:2014aa,:/content/aip/journal/apl/107/8/10.1063/1.4929727,AELM:AELM201500180,Yuan:2016aa}, among which creation by local spin-polarized current~\cite{PhysRevB.85.174416} and conversion from domain walls~\cite{Zhou:2014aa} have been experimentally realized~\cite{Romming636,Jiang283,Yu:2016aa} recently. The key idea shared among those proposals is to use spatially localized perturbations to generate skyrmions. Therefore, we can expect that spatially inhomogeneous perturbations from vortex beams is a way of generating a wider class of magnetic defects in chiral magnets reflecting the ring-shaped profile of those beams.

We model the effect of vortex beams on magnets as spatially nonuniform heating and study the spin dynamics on the basis of the stochastic Landau-Lifshitz-Gilbert (LLG) equation. We show that application of vortex beams to a two-dimensional film of chiral ferro- and antiferro- magnets can produce ring-shaped topological defects. In the simplest case, the defect is a bound state of a skyrmion and antiskyrmion and known as a $2\pi$ vortex~\cite{Bogdanov1999} (or skyrmionium~\cite{PhysRevLett.110.177205,PhysRevB.92.064412,PhysRevB.94.094420}) in literatures. Moreover, we will see that by changing beam parameters, we can also create defects with a multi-ring structure, namely $n\pi$ vortices (we call them {\it skyrmion multiplexes}) and conventional skyrmions. It is worth mentioning that our scheme using vortex beams is applicable to both FMs and AFMs, and to both metallic and insulating systems in the same manner.

The rest of this paper is organized as follows. In Sec.~\ref{sec. vortex} we review that vortex beams carry intrinsic OAM and have topological singularity with vanishing beam intensity at their center. We explain how the effect of vortex beams can be theoretically described in magnetic materials. Sections~\ref{sec. FM} and~\ref{sec. AFM} are devoted to our results. We demonstrate the emergence of a class of topologically stable magnetic defects in these systems other than ordinary skyrmions (and skyrmioniums). We determine an optimized way of applying vortex beams for creating these defects. Finally, we summarize our results in Sec.~\ref{sec: concl}.
\section{Vortex beams}\label{sec. vortex}
In Sec.~\ref{sec. vortex}A, we shortly review derivation of optical vortices, or Laguerre-Gaussian (LG) modes of Maxwell's equations in a vacuum based on the paraxial approximation. We explain that each eigenstate of the equations possesses intrinsic OAM. In Sec.~\ref{sec. vortex}B we consider how vortex beams interact with magnetic materials. Because of the mismatch in timescales between vortex beams and the dynamics of spins in chiral magnets, we can assume that the effect of vortex beams is heating which realizes nonuniform temperature proportional to the local beam intensity.

\subsection{Optical vortex, Laguerre-Gaussian solutions of Helmholz equation}\label{sec. 2a}
The dynamics of electromagnetic fields is described by Maxwell's equations. In particular, if we assume electromagnetic waves with fixed frequency $\omega$, their propagation in a vacuum is governed by the following
wave equations:
\begin{align}
\left(\Delta + \frac{\omega^2}{c^2}\right)\vec{E} = 0 \\
\left(\Delta + \frac{\omega^2}{c^2}\right)\vec{B} = 0,
\label{Max}
\end{align}
where $\Delta$ is a three-dimensional Laplacian and  $c$ is the speed of light in a vacuum.
These equations are equivalent to Helmholz-type differential equations.
\par
Vortex beams are defined by solutions of the wave equation
\begin{align}
(\Delta + k^{2})\psi(\vec{r}) = 0,
\label{eq: helmholz}
\end{align}
 in the cylindrical coordinate $(\rho, \phi, z)$, where $\rho$ is the radial coordinate, $\phi$ the azimuthal angle, and $z$ the coordinate along the cylindrical axis. In this coordinate, the Laplacian is written as $\Delta = \frac{\partial^{2}}{\partial \rho^{2}} + \frac{1}{\rho}\frac{\partial}{\partial \rho} + \frac{1}{\rho^{2}}\frac{\partial^{2}}{\partial \phi^{2}} + \frac{\partial^{2}}{\partial z^{2}}$. We assume that the beam propagates along the cylindrical axis. For example, the Maxwell's equation~\eqref{Max} for the electric field $\vec{E}(\vec{r})$ reduces to Eq.~\eqref{eq: helmholz} by setting$\frac{\omega^2}{c^2} = k^2$ and $\vec{E}(\vec{r}) = \vec{e}\psi(\vec{r})$ where $\vec{e}$ is the polarization vector. 
 By taking $\psi(\vec{r}) = u(\vec{r}) e^{i k z}$ we have 
\begin{align}
\left(\Delta_{T} +2 i k \frac{\partial}{\partial z} + \frac{\partial^2}{\partial z^2}\right)u(\vec{r}) = 0
\label{helm}
\end{align}
where $\Delta_{T} = \frac{\partial^{2}}{\partial r^{2}} + \frac{1}{\rho}\frac{\partial}{\partial \rho} + \frac{1}{\rho^{2}}\frac{\partial^{2}}{\partial \phi^{2}}$ is the transverse component of the Laplacian.
Let us employ paraxial approximation to Eq.~\eqref{helm}. Namely, we consider the case where $z$-dependence of $u(\vec{r})$ is small in the sense 
$|\frac{\partial^2 u}{\partial z^2}| \ll|\frac{\partial^2 u}{\partial x^2}|, |\frac{\partial^2 u}{\partial y^2}|$ and $|\frac{\partial^{2}u}{\partial z^{2}}| \ll 2k |\frac{\partial u}{\partial z}|$ 
where $k$ is the wavenumber in the $z$-direction. Then we can drop the second order derivative term $\frac{\partial^2 u}{\partial z^2}$ in Eq.~\eqref{helm} and have
\begin{align}
\left(\Delta_{T} + 2 i k \frac{\partial}{\partial z}\right)u(\vec{r}) = 0.
\label{parahelm}
\end{align}
Vortex beams, or LG modes~\cite{9780511795213} form a complete set of solutions for Eq.~\eqref{parahelm} and are given by 
\begin{align}
u^{LG}(\vec{r}) =\frac{1}{\sqrt{\left| w(z)\right| }}\left(\frac{\rho }{w(z)}\right)^{\left| m\right| } L_p^{\left| m\right| }\left(\frac{2 \rho ^2}{w(z)^2}\right) \times e^{-\frac{i k \rho ^2 z}{z^2+Z^2}}  e^{-i \chi (z) (\left| m\right| +2 p+1)}e^{i m \phi}e^{ -\frac{\rho
   ^2}{w(z)^2}},
   \label{eq: LG}
\end{align}
where the integers $p$ and $m$ label the mode and $L_p^{|m|}$ is the generalized Laguerre function. We note that by setting $p = m = 0$, Eq.~\eqref{eq: LG} falls into the usual Gaussian beam. The beam width $w(z) = w \sqrt{1 + \frac{|z|}{Z}}$ takes its minimum $w$, called the beam waist, at $z = 0$. Rayleigh range $Z$ is the distance from the focal plane along the propagation axis at which the cross-section of the beam becomes twice the minimum value. The phase factor determined by $\chi(z) =\tan^{-1}\left(\frac{z}{Z}\right)$ is called the Gouy phase. Since the OAM around the propagation axis is given by the eigenvalue of $L^{z} = -i\hbar\partial/\partial\phi$, the phase twist $e^{i m \phi}$ in Eq.~\eqref{eq: LG} yields OAM $\hbar m$. This factor requires the topological singularity $u^{LG}(0,\phi,z) = 0$ to maintain the single-valuedness of electromagnetic fields at the origin $\rho = 0$. Hence the OAM leads to ``doughnut-shaped" transverse intensity profile of the propagating vortex beams. We emphasize that the OAM is a property of $u(\vec{r})$ or $\psi(\vec{r})$ and has nothing to do with the polarization vector $\vec{e}$ of the electromagnetic waves which corresponds to the spin degrees of freedom of photons.\par
Compared to the optical vortex, vortex beams of electrons are more subtle. Here we just mention that in a particular setup, the wave function of electrons also acquires the factor $e^{i m \phi}$ with $m$ an integer (for details, see Ref.~\cite{PhysRevLett.99.190404}).

\subsection{Effect of vortex beams on chiral magnets}
\begin{figure}[htbp]
   \centering
\includegraphics[width = 165mm]{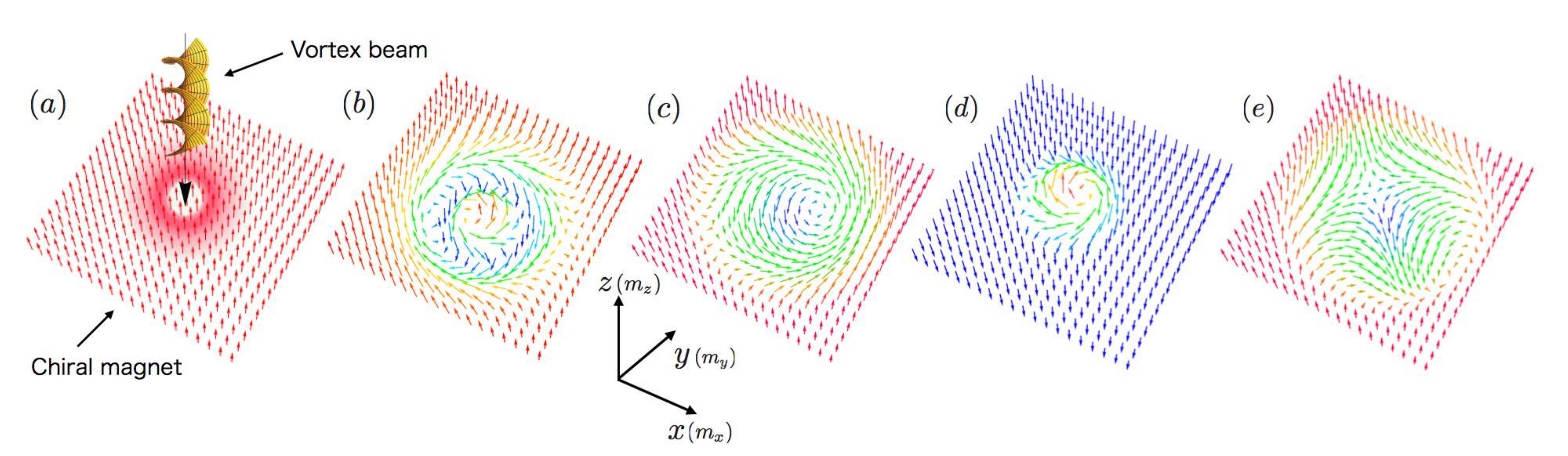}
       \caption{(a) Schematics of our setup. The vortex beam is irradiated to a chiral magnet and induces nonuniform temperature proportional to the local intensity of the beam. Arrows represent spins in chiral magnets [see Eq.~\eqref{model}] and the colors represent their $z$ component (red for $+1$, blue for $-1$, and green for $0$). (b) A topological defect created by the vortex beam, skyrmionium. (c),(d) Spin texture of skyrmions in ferromagnetic backgrounds with up and down spins. (e) Spin texture of an anti-skyrmion.}
          \label{schematics}  
          \end{figure}

\begin{figure}[htbp]
   \centering
\includegraphics[width = 130mm]{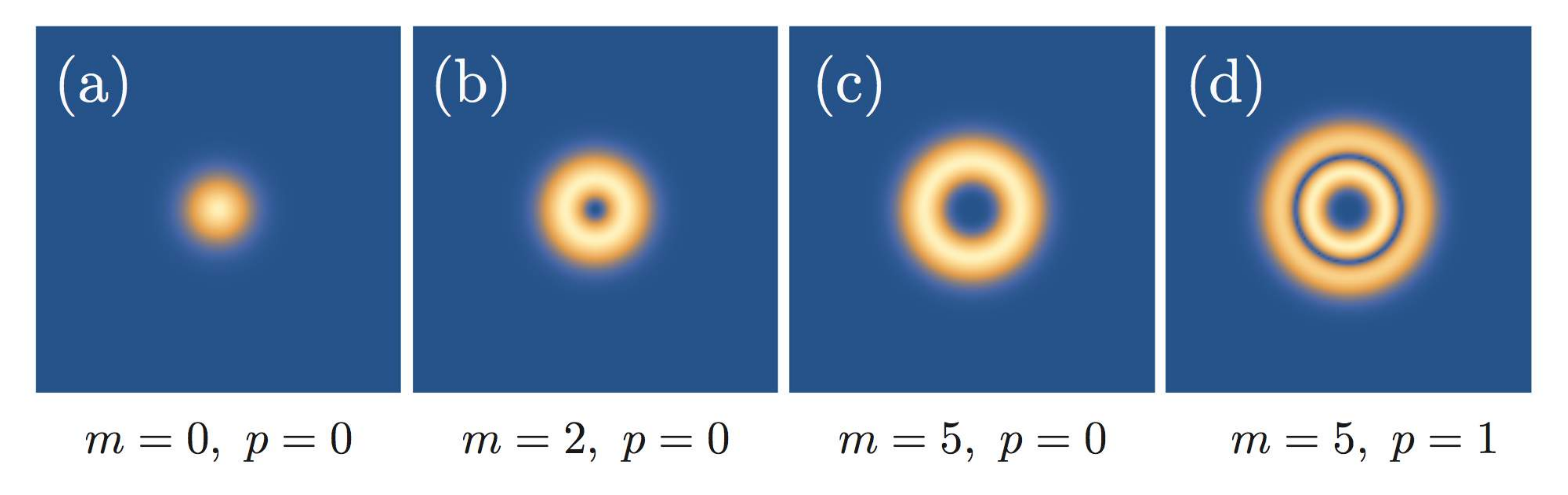}
       \caption{Intensity of Laguerre-Gaussian modes at the focal plane $|u^{LG}(\rho, \phi, z = 0)|^2$ for several choices of $p$ and $m$. The brightness is proportional to the intensity. For non-vanishing values of $m$, the intensity at the center is exactly zero and there appear $(p + 1)$-fold rings in the intensity distribution. We also see that for larger $m$, the size of rings becomes larger with increasing $m$. }
          \label{beam pattern}  
          \end{figure}

 In Fig.~\ref{schematics}(a) we show the schematics of our setup. We place a two-dimensional film of chiral magnets (the figure is for chiral FMs) at the focal plane ($z = 0$) of the vortex beams with a ring-shaped spatial profile:
\begin{align}
u^{LG}(\rho, \phi,0) = \frac{\left(\frac{\rho }{w}\right)^{\left| m\right| } e^{-\frac{\rho ^2}{w^2}+i m \phi } L_p^{\left| m\right| }\left(\frac{2 \rho ^2}{w^2}\right)}{\sqrt{\left| w\right| }}.
\label{LGmode}
\end{align}
In Fig.~\ref{beam pattern}, for several choices of $p$ and $m$, we show the beam intensity of the vortex beams $|u^{LG}(\rho, \phi, 0)|^2$ at $z = 0$. For $m = 0$ the beam~\eqref{LGmode} reduces to the usual Gaussian beam [Fig.~\ref{beam pattern}(a)] whose intensity peaks at the center ($\rho = 0$). For $m\neq0$ as we show in Fig.~\ref{beam pattern}(b, c, d), we have topological singularities at the center and the intensity distribution looks like $(p+1)$-fold rings. Such a multiring structure can be also realized by superimposing several single-ring vortex beams with different ring sizes.

 \par


The length scale of magnetic textures in chiral magnets is determined by the exchange coupling and the strength of DM interaction. In particular, the size of skyrmions in them is specified by the ratio of the exchange and DM coupling. If there can exist other kinds of stable defects induced by the vortex beams, their size should not be so much different from that of skyrmions. This naive expectation is numerically verified in Sec.~\ref{sec. FM}. Then, taking into account the diffraction limit of lasers, we expect that the wavelength of the vortex beams suitable for our purpose is  close to the size of the skyrmions in the target material. The typical size of the skyrmions in chiral FMs is O$(1)$ to O$(10)$ nm, so that the proper wavelength is in the region of extreme ultraviolet (EUV) lights.  We note that by using proper materials with larger skyrmions with a diameter several hundreds of nano meters to a few micro meters~\cite{ShibataK.:2013aa, Jiang283,Yu_2016,largesk,Woo:2016aa,Moreau-LuchaireC.:2016aa,Boulle:2016aa}, the natural wavelength can be that of visible lights. In this paper we assume parameters of typical chiral FMs. Optical vortices with EUV wavelength can be generated in the same way as that of visible lights, by using a nonlinear optical effect or holography~\cite{PhysRevLett.111.083602,Terhalle:11}. As for electron vortex beams, it is much easier to focus them to this wavelength. \par

If the wavelength is about $10$ nm, the frequency of the optical vortex is about $30$ PHz. This is much faster than the dynamics of spins in magnetic materials, whose timescale is typically of the order of THz. Therefore, the optical vortex with this frequency cannot induce any coherent dynamics of spins as it is. Instead, vortex beams create hot electrons and excite high-energy lattice vibrations which (locally) equilibrate within $100$ fs to $1$ ps~\cite{RevModPhys.82.2731} (EUV lights are easily absorbed by matter). We note that there are several non-thermal effects of high-frequency lasers on magnets mediated by electronic degrees of freedom such as Raman effect, Faraday/Kerr effect, and Cotton-Mouton effect~\cite{RevModPhys.79.175,Lemmens:2003aa,PhysRev.143.574}. For example, in Ref.~\cite{Bigot:2009ab}, coherent interaction between spins in ferromagnets and a femto second laser pulse is observed, behind which the conservation of spin angular momentum plays an essential role.  In this work, we focus only on the heating effect and ignore these non-thermal spin-photon couplings, employing simple spin models of chiral magnets for numerical calculations.

That is, we assume that vortex beams only raise the local temperature of the system in accord with the beam intensity. We note that in reality, the temperature gradient achieved in experiments may not be so clear. If we use materials with sufficiently large skyrmions, there would be no problems, but if the length scale is in the EUV region (wavelength $\sim 10$ nm), some modifications would be required. Nevertheless, we expect that for our purpose, printing the spatial profile of vortex beams to chiral magnets, details of the temperature distribution is unimportant, as long as the realized temperature reflects the ring-shaped spatial profile of vortex beams. Indeed, we numerically verified that vertex beams with various OAM integer $m$ show qualitatively the same outcomes in our setup. Therefore, in this work we assume that vortex beams simply induce temperature proportional to the local beam intensity: $T(\vec{r}) \propto |u^{LG}(\rho, \phi, 0)|^2$~\footnote{If we take the electric field to be a LG mode with $\vec{e}_{p}\perp \hat{z}$, Faraday's law requires the magnetic field to have very small $z$ component, which can be nonzero even at $\rho = 0$. Here for simplicity we ignore the heating caused by that small magnetic field.} and see whether the heating can generate defects with unconventional spatial profile. We mention that for chiral FMs, it is theoretically predicted that local but uniform heating can lead to ordinary skyrmions~\cite{Koshibae:2014aa}.

The situation for electron vortex beams is more subtle because electrons are charged and the beams generate electromagnetic fields which can affect spin structures of chiral magnets. However, the intensity of electron vortex beams generated by scanning or transmission electron microscopy (TEM) equipments is generically small at present so that the dominant effect will be heating just as optical vortices. 

Here we shortly comment on the use of THz optical vortex with which we can coherently control the dynamics of spins by using the direct coupling between spins and electromagnetic fields, not by heating. 
Due to the large length scale ($300$ $\mu$m for $1$ THz) and the diffraction limit, naively we expect that for THz optical vortices as they are, possible manipulations would be limited to large-scale structures such as magnetic domain walls.
However, recently it is becoming possible to break the diffraction limit of lights with plasmonics techniques~\cite{Heeres:2014aa,PhysRevX.2.031010, tanaka}. With such techniques, we can explore non-thermal control of nano-scale magnetic textures in solids with THz optical vortices~\cite{inprep}. Nano-focusing with plasmonics can be also useful for heating. Usually, visible lights are much more accessible than EUV lights for experimentalists. Concerning practical applications therefore, combining visible optical vortex and plasmonics would be more promising than EUV optical vortex.
 
\section{Chiral ferromagnet}\label{sec. FM}
In this section, we discuss topological defects induced by vortex beams in two-dimensional film of chiral FMs. First we shortly review static properties of chiral FMs, the phase diagram and characteristic magnetic defects, skyrmions in a canonical model of them. Then, we explain our numerical calculations based on (stochastic-) Landau-Lifshitz-Gilbert equation. We demostrate the generation of a class of ring-shaped topological defects by vortex beams and optimize the way of applying the beams.

 
  \begin{figure}[htbp]
   \centering
\includegraphics[width = 100mm]{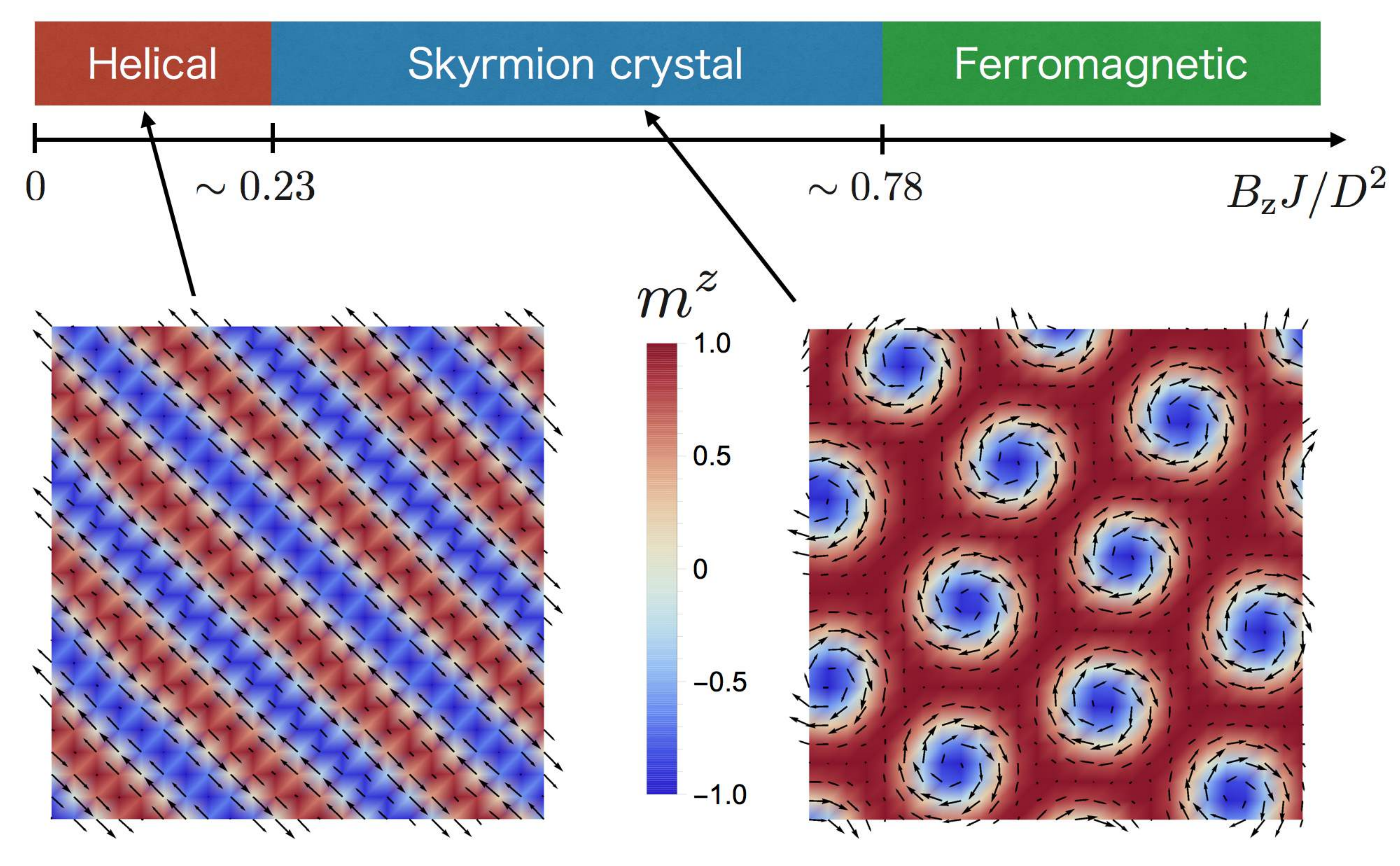}
       \caption{Ground state phase diagram of the canonical model of chiral ferromagnets~\eqref{model} (reproduced from Ref.~\cite{PhysRevLett.108.017601}). There appear three distinct phases depending on the magnitude of the external magnetic field applied in the $z$-direction. For the helical order phase and skyrmion crystal phase, we give their spin texture obtained by numerical simulations based on the LLG equation, using arrows and colors.}
          \label{phasediag_chiralFM} 
  \end{figure}  
 

\subsection{Static properties of chiral magnets}
Chiral FMs such as B20-type compounds MnSi~\cite{PhysRevB.70.075114,Muhlbauer:2009aa,PhysRevLett.102.186602,PhysRevLett.102.197202}, $\mathrm{Fe}_{1-x}\mathrm{Co}_x$Si~\cite{Yu:2010aa, PhysRevB.81.041203}, or FeGe~\cite{Yu:2011aa,PhysRevLett.108.267201,Woo:2016aa}, are characterized by their relatively large DM interaction. A canonical model of chiral FMs is the following two-dimensional classical spin model defined on a square lattice with $\vec{m}_{\vec{r}}$ being the spin at the site $\vec{r}$~\footnote{Precisely speaking, by using the electron spin $\vec{s}$ the magnetic moment $\vec{m}$ is defined as $\vec{m} = - \vec{s}$. }:
\begin{align}
H = - J \sum_{\vec{r}} \vec{m}_{\vec{r}}\cdot\left(\vec{m}_{\vec{r} + a \vec{e}_{x}}+ \vec{m}_{\vec{r} + a \vec{e}_{y}} \right) 
+\sum_{\vec{r}}\vec{D}_{i}\cdot\left(	\vec{m}_{\vec{r}}\times \vec{m}_{\vec{r}+a\vec{e}_{i}}		 \right)
-B_z\sum_{\vec{r}}m^{z}_{\vec{r}},
\label{model}
\end{align}
where $a$ is the lattice constant and $\vec{e}_{i}$ is the unit vector along the $i$ axis ($i = x, y$). The exchange coupling $J > 0$ represents ferromagnetic Heisenberg interaction and $\vec{D}_{i} $ is DM vector on the bond $(\vec{r}, \vec{r}+a\vec{e}_{i})$, and $B_{z}$ is the external magnetic field applied in the $z$-direction. Hereafter we normalize the length of $\vec{m}_{\vec{r}}$ to unity. As long as we are interested in physics like spin waves or magnetic defects whose length scale is much longer than the lattice constant, we expect that the model~\eqref{model} works as a standard model of chiral FMs. Indeed, many experimental results on real materials are well described by this model (see for example Ref.~\cite{Seki_BOOK}), even though their microscopic Hamiltonians must be much more intricate. \par

What is peculiar about chiral magnets is the emergence of topologically stable magnetic defects, skyrmions. In Fig.~\ref{schematics}(c, d), we show skyrmions for a particular choice of DM vectors $\vec{D}_{i} = D \vec{e}_{i}$ ($D>0$) and the external magnetic field $\vec{B} = \pm B_{z}\vec{e}_{z} $. In Fig.~\ref{schematics}(e) we also show the spin texture of a so-called anti-skyrmion. (Anti-) skyrmions are characterized by their non-vanishing skyrmion number:
\begin{align}
N_{\mathrm{SK}} = \frac{1}{4\pi}\int \vec{m}_{\vec{r}}\cdot \left(\frac{\partial \vec{m}_{\vec{r}}}{\partial x}\times \frac{\partial \vec{m}_{\vec{r}}}{\partial y} \right)d^{2}r.
\label{skyrmion_num}
\end{align}
In a continuous space, $N_{\mathrm{SK}}$ must be quantized to an integer, so that (anti-) skyrmions are stable against any continuous deformations. Even for lattice systems, where the quantization is incomplete, skyrmions and anti-skyrmions are topologically stable if their size is much larger than the lattice constant. Nevertheless, in chiral FMs, anti-skyrmions are energetically unstable~\cite{Koshibae:2016aa} and have short lifetime for the given DM vectors $\vec{D}_{i} = D \vec{e}_{i}$. \par

The phase diagram of the model is well studied (see for example Ref.~\cite{Hubert1994}). For a fixed exchange and DM interaction, when the external magnetic field is very small, the ground state develops a helical magnetic order with a single wavelength determined by the magnitude of the DM interaction. As we increase the magnetic field, the helical order becomes unstable and the ground state turns into a triangular lattice of skyrmions. For very large magnetic field, spins become polarized in the direction of the field and the ferromagnetic ground state is stabilized. Therefore, the phase diagram of this model is summarized as Fig.~\ref{phasediag_chiralFM}. We present the critical values of the magnetic field and spin textures of typical states in the helical order phase and skyrmion crystal phase. In addition to the skyrmion crystal phase, skyrmions can appear as energetically stable isolated magnetic defects in the ferromagnetic phase. 
We note that anisotropy energy, which lacks in the model~\eqref{model}, can also stabilize skyrmions. Later, we will discuss the situation for chiral AFMs, where we introduce anisotropy energy for the stabilization of skyrmions in them. 
 

 \begin{figure}[htbp]
   \centering
\includegraphics[width = 150mm]{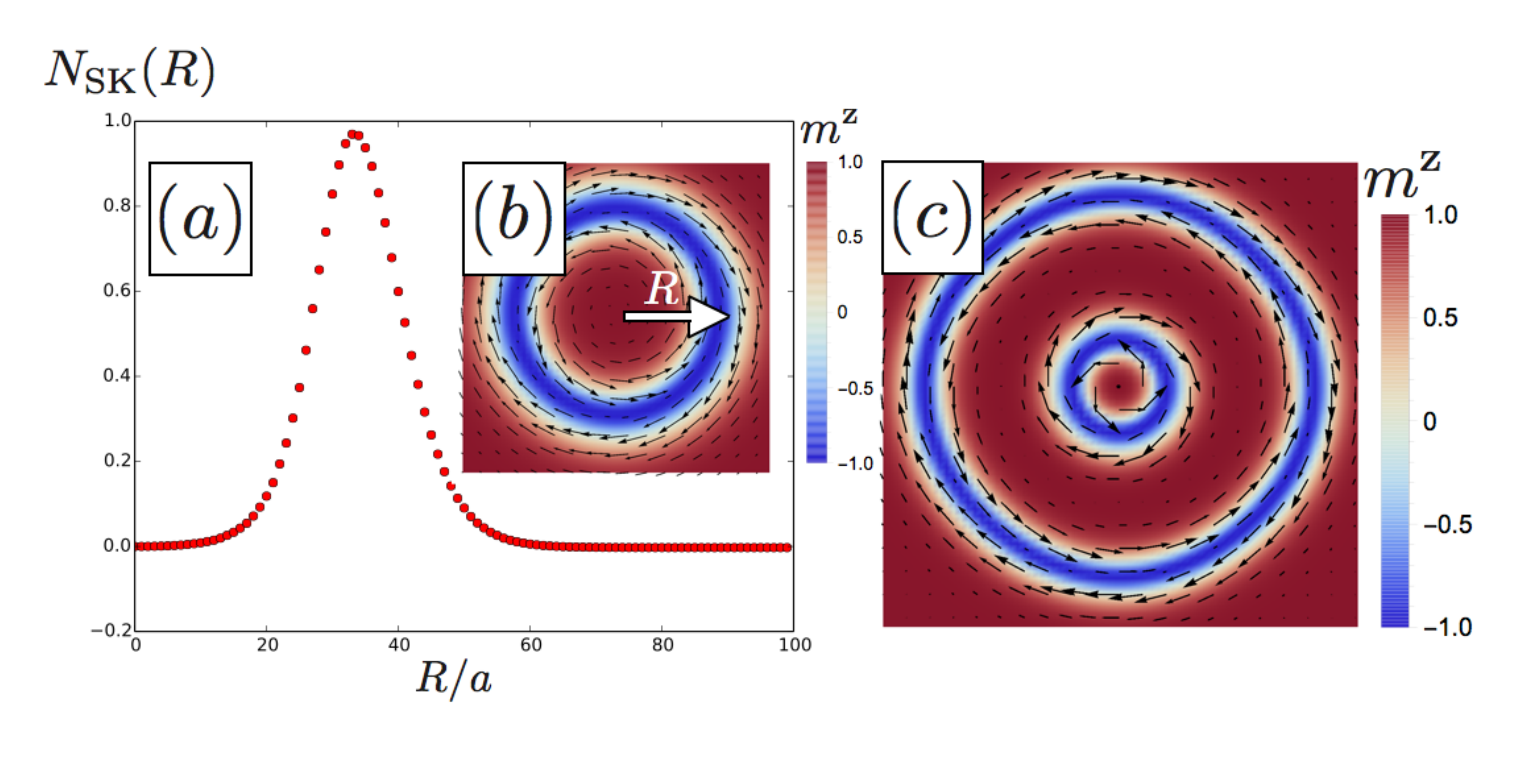}
\caption{Skyrmion duplex and quadplex obtained by numerical simulation of the model~\eqref{model} for $J = 1$, $D=0.15$, and $B_{\mathrm{z}}=0.014$. (a) Cumulative skyrmion number $N_{\mathrm{SK}}(R)$ of the skyrmion duplex. as a function of $R/a$, where $a$ is the lattice constant. The skyrmion duplex, or skyrmionium is a skyrmion with positive $N_{\mathrm{SK}}$ [Fig.~\ref{schematics}(d)] surrounded by that with negative $N_{\mathrm{SK}}$ [Fig.~\ref{schematics}(c)]. (b) Spin texture of the skyrmion duplex with diameter about 50 sites. (c) Spin texture of the skyrmion quadplex whose whole size is approximately 160 sites in diameter. The arrows represent the in-plane components of spins and the color indicates their $z$ component. 
}
\label{SkD combined}          
    \end{figure} 
    

\subsection{Skyrmion multiplex}
With numerical calculations based on stochastic LLG (sLLG) equation, we show that the ring-shaped temperature profile induced by vortex beams offers a way to create ring-shaped topological magnetic defects shown in Fig.~\ref{schematics}(b) and Fig.~\ref{SkD combined}(b)(c). Details of numerical methods and optimal way of applying vortex beams will be discussed in Sec.~\ref{sec: FM}.

To see the topological nature of these defects, in Fig.~\ref{SkD combined}(a), we present the cumulative skyrmion number for the defect Fig.~\ref{SkD combined}(b)
\begin{align}
N_{\mathrm{SK}}(R) = \frac{1}{4\pi}\int_{r<R} \vec{m}_{\vec{r}}\cdot \left(\frac{\partial \vec{m}_{\vec{r}}}{\partial x}\times \frac{\partial \vec{m}_{\vec{r}}}{\partial y} \right)d^{2}r. 
\label{cumulative}
\end{align}
The integration is performed within the circle with radius $R$ measured from the center of the defect. The spin texture and the $R$ dependence of $N_{\mathrm{SK}}(R) $ clearly show that the ring-shaped defect is a bound state of two skyrmions in Fig.~\ref{schematics}$(c)(d)$ where the former (latter) has $N_{\mathrm{SK}}=-1$ (+1) if placed in a continuous space and isolated. Namely, its spin configuration outside a certain radius, at which $m^{z} = -1$, is that of the skyrmion in Fig.~\ref{schematics}(c) and the spins inside that radius form the skyrmion in Fig.~\ref{schematics}(d). This kind of a ``dounut-shaped" defect is known as $2\pi$ vortex~\cite{Bogdanov1999} (also called as skyrmionium recently~\cite{PhysRevLett.110.177205,PhysRevB.92.064412,PhysRevB.94.094420}), but we find that the skyrmionium is just the simplest case of a family of possible topological defects with ring-shaped structure induced by vortex beams. In general, by changing the spatial structure of the beams by for example, using the different radial index $p$, we can create general multi-ring structures, namely $n\pi$ vortex. In this paper, we call skyrmionium as skyrmion duplex (SkD) and those generic ring-shaped structures as skyrmion multiplexes.  In Fig.~\ref{SkD combined}(b, c), we show the spin structure of a skyrmion duplex (skyrmionium) and skyrmion quadplex ($4\pi$ vortex). Although both of them are excited states within the skyrmion crystal phase, we find that they are, once formed, stable within the framework of our LLG calculations. We note that recently, their stability in a nanodisk geometry was systematically studied in Ref.~\cite{PhysRevB.94.094420}. If those topological defects actually have sufficiently long lifetime in real materials as theoretically predicted, their spin structures can be experimentally observed with Lorentz TEM, just as done for soliton lattices and skyrmion lattices~\cite{Yu:2010aa,PhysRevLett.108.107202}. 
 
\subsection{Formation of skyrmion multiplex}
We numerically calculate the time-evolution of spins based on the sLLG equation~\cite{PhysRevB.58.14937,Koshibae:2014aa} for the model~\eqref{model}:
\begin{align}
\frac{d\vec{M}_{\vec{r}}}{dt} &= - \gamma \vec{M}_{\vec{r}} \times \left(-\frac{\partial H}{\partial \vec{M}_{\vec{r}}} + \vec{h}_{T(\vec{r})}(t)\right)+ \alpha\frac{\vec{M}_{\vec{r}}}{|\vec{M}_{\vec{r}}|}\times \frac{d\vec{M}_{\vec{r}}}{dt},
\label{sLLG}
\end{align} 
where $\vec{M}_{\vec{r}} = \hbar \gamma \vec{m}_{\vec{r}}$ and $\gamma$ is the gyromagnetic ratio. The time coordinate is $t$. The second term in the right hand side in Eq.~\eqref{sLLG} is the so-called Gilbert damping term which describes dissipation. The dimensionless constant $\alpha$ characterizes the strength of the dissipation. In the framework of the sLLG equation, the effect of heating is treated as a random field $\vec{h}_{T}(t)$ satisfying
\begin{align}
\Braket{ h_{T}^{ \mu}(t)} &= 0, \nonumber \\
\Braket{ h_{T(\vec{r})}^{\mu}(t)h_{T(\vec{r'})}^{\nu}(t')} &= \sigma(\vec{r}) \delta^{\mu, \nu}\delta(\vec{r}-\vec{r'})\delta(t-t'),
\label{random} 
\end{align}
where $\mu, \nu = x, y, z$. Here $\sigma(\vec{r})$ is determined by temperature at $\vec{r}$ from the fluctuation-dissipation theorem: $\sigma(\vec{r}) = 2 k_{B}T(\vec{r})\alpha/(\gamma^2\hbar)$. Therefore, by using Eq.~\eqref{sLLG} with this random field we can numerically simulate the time evolution of the model~\eqref{model} at finite temperatures. We use the Heun method with the time step $\Delta t = 0.02$ for numerical integration of the sLLG equation with the linearization technique~\cite{Seki_BOOK}.
\par
We fix the parameters as $J = 1$, $D = 0.15$, and $\alpha = 0.1$. Although we focus on the skyrmion crystal phase but the initial state of simulations ($t = 0$) is taken as a metastable, perfect ferromagnetic state i.e. $\vec{m}_{\vec{r}} = (0, 0, 1)$ for all $\vec{r}$. For these values of  $J$ and $D$, the phase boundary between the helical order phase and the skyrmion crystal phase is $B_z = 0.0052$ and that between the skyrmion crystal and the ferromagnetic phase is $B_z = 0.018$.
After several trials and errors we find that ``annealing" the spins is advisable to obtain topological defects. Hence we fix the time dependence of the vortex-beam driven temperature as $T(t) = T_0 \left( 1 - \frac{t}{t_0}\right) \Theta(t) \Theta(t_{0}-t)$ with $t_0 = 500$. Here $\Theta(x)$ is the step function. Namely, assume that the temperature is instantaneously raised to its maximum ($T_{0})$ and gradually cooled down. The time is measured in the unit of $\hbar/ J$, which is estimated to be $0.7$ ps for $J = 1$ meV. For this strength of $J$, $B_z = 0.01$ corresponds to $0.173$ Tesla and $T = 1$ does $11.6$ K. As we mentioned in the previous section, we assume that the beam-induced temperature is proportional to the local beam intensity. Therefore, we have $T(t, \vec{r}) = T(t) \left[|u^{LG}(\rho,\phi, 0)|^2/\underset{\vec{r}}{\mathrm{max}}(|u^{LG}(\rho,\phi, 0)|^2)\right]$.

\begin{figure}[htbp]
   \centering
\includegraphics[width = 140mm]{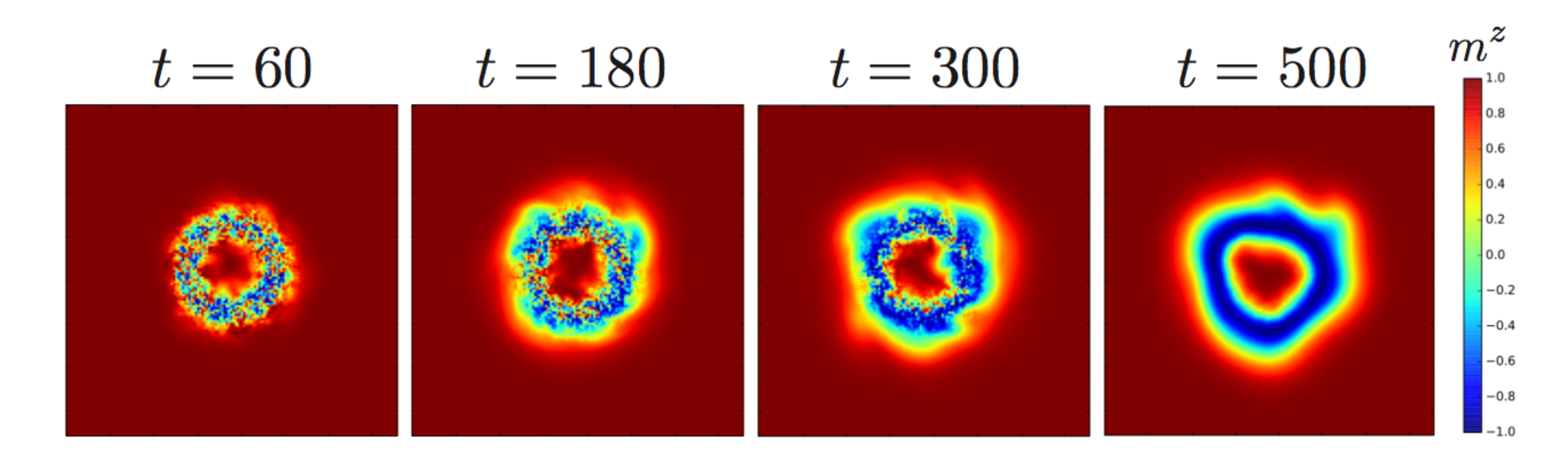}
       \caption{Time evolution of the $z$ component of spins for a particular trial with parameters $B_z = 0.01, D = 0.15, w = 12.5 a, t_0 = 500$, and $T_{0} = 2$. We start from the metastable ferromagnetic state at $t = 0$. At $t = 0$ we suddenly raise the temperature in accord with the intensity of the vortex beam with $p =0$ and $m=5$. Then the temperature is lowered gradually and finally, a skyrmion duplex is formed. The time is measured in the unit of $\hbar/(J)$. When $J = 1$ meV the time unit corresponds to $0.7$ ps. The system consists of 150 $\times$ 150 sites and a periodic boundary condition is imposed.}
          \label{FM_SkD_formation} 
  \end{figure}

\begin{figure}[htbp]
   \centering
\includegraphics[width = 140mm]{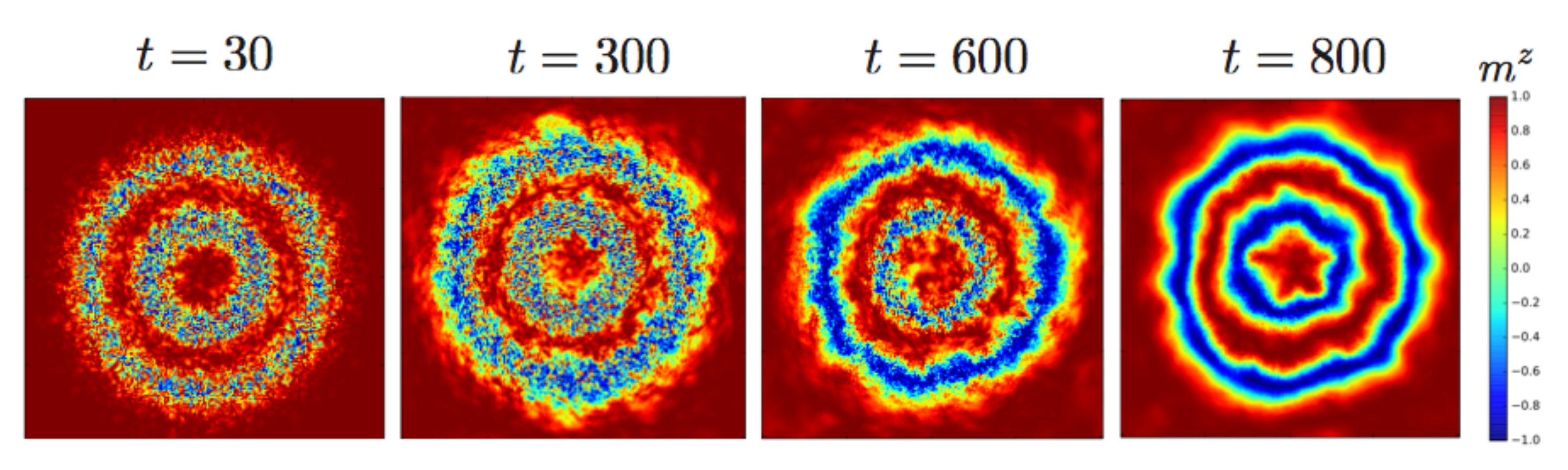}
       \caption{Time evolution of the $z$ component of spins for a particular trial with parameters $B_z = 0.011, D = 0.15, \alpha = 0.1, w = 100a/3$, $T_{0} = 4$, and $t_{0} = 800$. For $p = 1$ and $m = 3$, the intensity of the beam takes the form of a double-ring and a skyrmion quadplex is formed after the irradiation. The system consists of 200 $\times$ 200 sites and a periodic boundary condition is imposed.}
          \label{triplex_process} 
  \end{figure}

\begin{figure}[htbp]
   \centering
\includegraphics[width = 140mm]{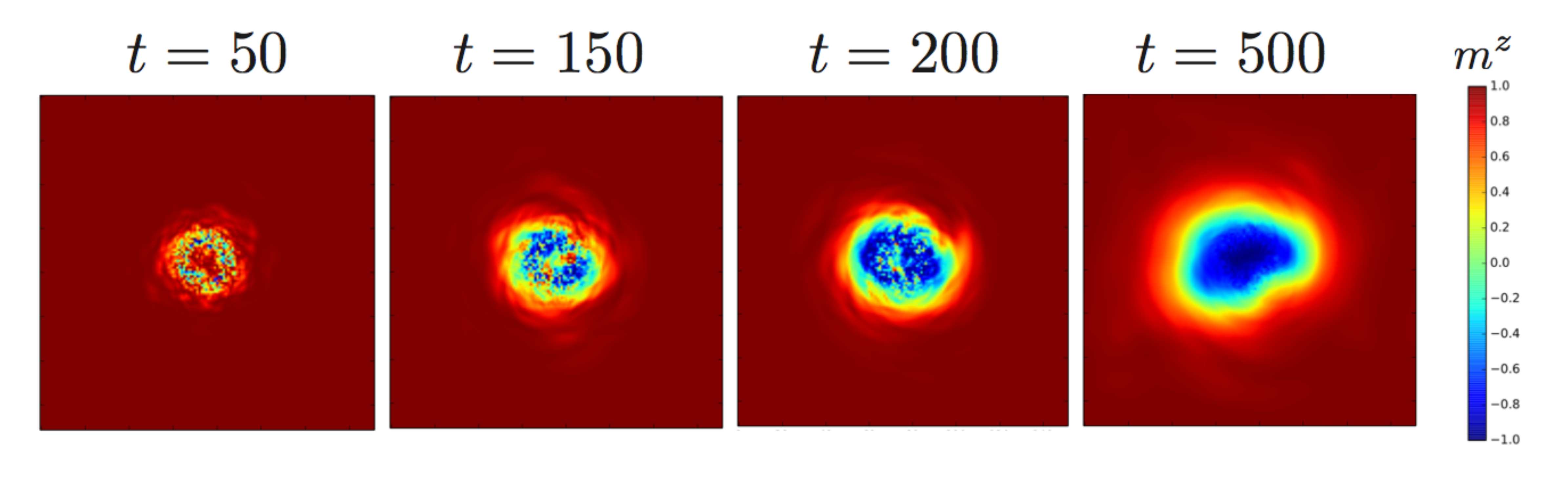}
       \caption{Time evolution of the $z$ component of spins for a particular trial with parameters $B_z = 0.01, D = 0.15$, $t_{0} = 500$, and $T_{0} = 1.5$. We still use a vortex beam with $p = 0$ and $m = 5$, but the beam waist is set to be small: $w = 6 a$.  Due to the small beam waist, the topological singularity in the temperature profile is smeared and as a result, an ordinary skyrmion is created. The system consists of 150 $\times$ 150 sites and aperiodic boundary condition is imposed. }
          \label{failed} 
  \end{figure}

In Figs.~\ref{FM_SkD_formation}, \ref{triplex_process}, and \ref{failed}, we present typical time evolutions of the magnetic texture under the annealing processes. In Fig.~\ref{FM_SkD_formation} we present the result for $p = 0$ and $m = 5$, namely, for a single-ring beam profile. We see that the spin texture perturbed by the vortex beam is annealed to form an SkD. In Fig.~\ref{triplex_process}, we show the time evolution for $p = 1$, $m = 3$. When $p = 1$, as we saw in Fig.~\ref{beam pattern}(d), the intensity profile of the beam looks like doublerings and we see that its spatial structure is successfully printed to spins as a skyrmion quadplex, which consists of two antiskyrmions and two skyrmions. When the beam waist is small, the topological singularity of vortex beams becomes unimportant and our setup becomes similar to the situation of Ref.~\cite{Koshibae:2014aa}, where uniform heating caused by a nearfield is proposed as a way of creating ordinary skyrmions. Indeed, as shown in Fig.~\ref{failed}, using vortex beams with a small beam waist, we can create ordinary skyrmions for $p = 0$ and $m = 5$, the same as Fig.~\ref{FM_SkD_formation}. Therefore, by changing the beam waist, both a skyrmion ($\pi$ vortex) and SkD can be generated using the same method. These results indicate that vortex beams with proper beam waist and the integer $p$ yield $n\pi$ vortices with an arbitrary integer $n$ as long as they are allowed as energetically stable defects in the target material.

\subsection{Optimal way of applying vortex beams}\label{sec: FM}
\begin{figure}[htbp]
   \centering
\includegraphics[width = 120mm]{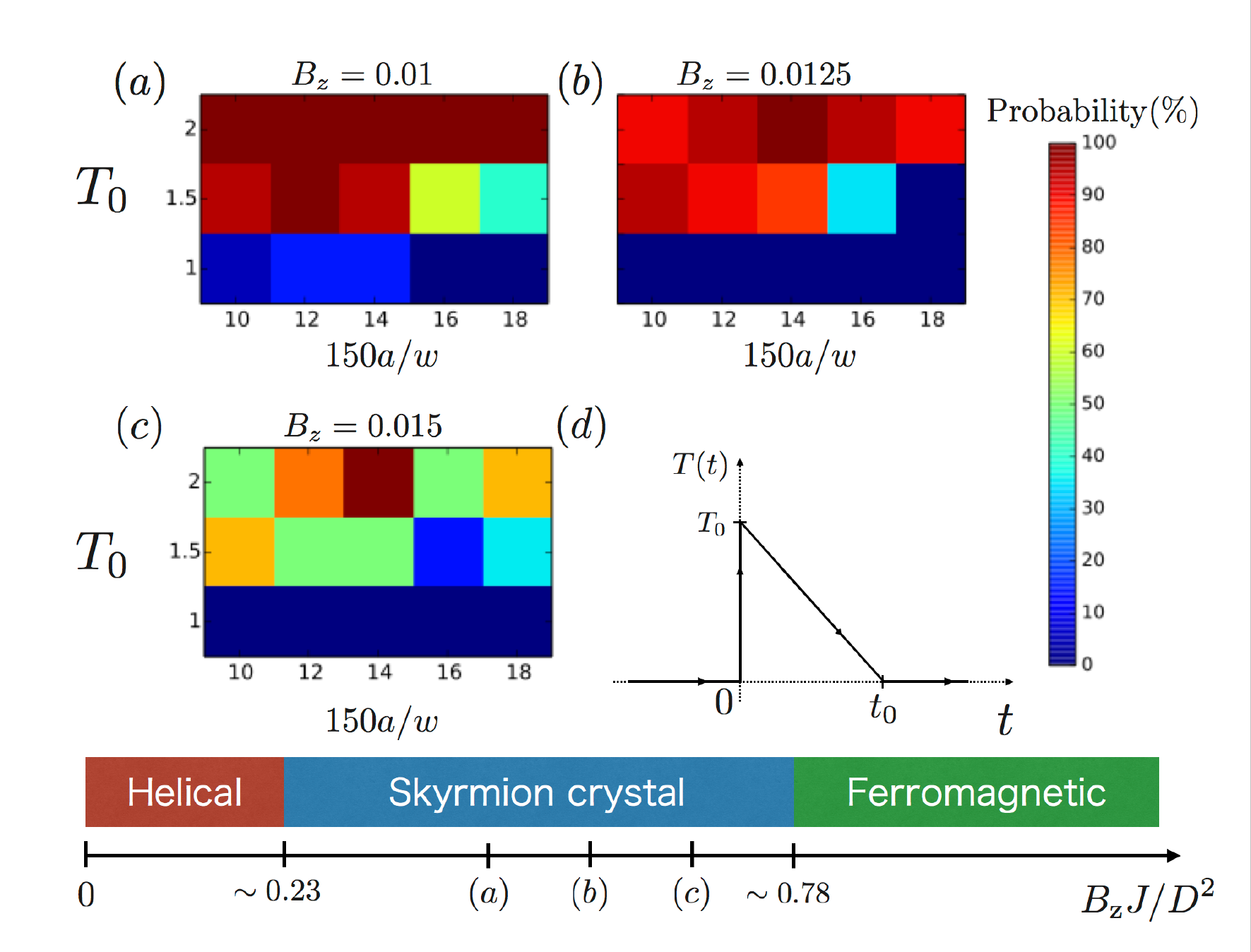}
       \caption{Success probability of creating skyrmion duplexes by vortex beams with $p = 0$ and $m = 5$, for $J = 1$, $D = 0.15$, and $\alpha = 0.1$. (a) Probability for $B_z = 0.01$, (b) $B_z = 0.0125$, and (c) $B_z = 0.015$. The local temperature $T(t, \vec{r})$ is set to be proportional to the intensity of the beam: $T(t, \vec{r}) = T(t) \left[|u^{LG}(\rho,\phi, 0)|^2/\mathrm{max}(|u^{LG}(\rho,\phi, 0)|^2)\right]$. Here the time dependence of the temperatures is given as panel (d): $T(t) = T_0 \left( 1 - \frac{t}{t_0}\right) \Theta(t) \Theta(t_{0}-t)$ with $t_{0} = 500$. The highest probability is achieved when the magnetic field is small and the beam waist satisfies $w \sim 11.5 a$ for which the wavelength of the beam is comparable with the size of skyrmions. We also show where the chosen magnetic fields $B_z$ locate in the phase diagram.}
          \label{phasediag_alpha=0.1} 
  \end{figure}

Here we determine the optimal way of applying vortex beams to achieve a high probability of creating SkDs with single-ring vortex beams ($p = 0$) by changing other parameters. We try the simulations in a periodic system with $150\times150$ sites and calculate the success probability of creating SkDs. We take the peak temperatures $T_0= 1, 1.5,$ and $2$ and the time dependence of the temperature as shown in Fig.~\ref{phasediag_alpha=0.1}(d). In Fig.~\ref{phasediag_alpha=0.1} we show the success probability of creating SkDs, $(N/20\times 100)$ $\%$, where $N$ is the number of success trials in which we obtain SkDs, and 20 is the total number of trials for each set of $(T_0, B_z, w)$. The probability tops when the magnetic field is small and the beam waist satisfies $150 a \sim13 w$. If we take $w = 0.5\lambda$ and $a = 5 \mathrm{\AA}$ as the optimum wavelength we have $\lambda \sim 12$ nm, which is close to that of EUV lasers and to the size of skyrmions for the given parameters. We can also use electron vortex beams to achieve this wavelength. The optimal wavelength would be generically determined by the size of skyrmions. Therefore for chiral ferromagnetic films with large skyrmions like CoFeB/Ta~\cite{Jiang283,Yu_2016}, optical vortices of visible lights would be appropriate. \par

We numerically confirm that the size of SkDs grows with decreasing $B_z$ and it is impossible to create stable SkDs in the ferromagnetic phase of the present model. As we decrease the magnetic field $B_z$, the ferromagnetic initial state we assume would become more fragile in reality, while the SkDs themselves are easier to be created and are energetically stabler. 
 
We note that creation of SkDs using an electric current pulse with a period of a few hundred pico-seconds was discussed recently in Ref.~\cite{PhysRevB.94.094420}. In their case, the timescale of the creation is a few hundred pico seconds for $J \simeq 94$ meV. For this strength of $J$, the time unit is $\hbar/J = 7.5$ fs, and $t_{0} = 500$ corresponds to $3.8$ ps, so that our scheme using vortex beams is 100 times faster than the method with a current pulse, though our assumption for local equilibration no longer holds in this timescale. Moreover, their scheme using current pulse cannot induce multi-ring defects like a quadplex and cannot be applied to insulating materials unlike our scheme.

Finally, we comment on the recent experimental study which observes various intricate magnetic defects including skyrmioniums in a ferrimagnetic film~\cite{PhysRevLett.110.177205}. In this study, a simple laser pulse without OAM is used to cause ultrafast magnetization reversal to the system. Due to the strong anisotropy energy in the direction perpendicular to the film, this system allows skyrmions to appear even in the absence of an external magnetic field. In this case, the single-domain ferromagnetic state is less stable than our model where the external magnetic field explicitly breaks the degeneracy between two ferromagnetic states aligned along the field direction. Hence, by perturbing the spin structure sufficiently, there can appear various magnetic defects, though the species of the defects created in that way are not well controlled as opposed to those by our scheme.
 
\section{Chiral antiferromagnet}\label{sec. AFM}
Next we move onto chiral AFMs. The basic idea is the same as the ferromagnetic case. Namely, we consider printing the spatial structure of vortex beams to two-dimensional film of chiral AFMs as topological defects. We model the application of vortex beams as heating which realizes temperature proportional to the local intensity of the beams, and examine the time evolution of spins by the sLLG equation. 

We first review the phase diagram of a canonical model of chiral AFMs. Then we show the emergence of an antiferromagnetic analog of skyrmion multiplexes after the laser irradiation and discuss how to achieve a high success probability of their creation. 

\subsection{Static properties} 
\begin{figure}[htbp]
   \centering
\includegraphics[width = 120mm]{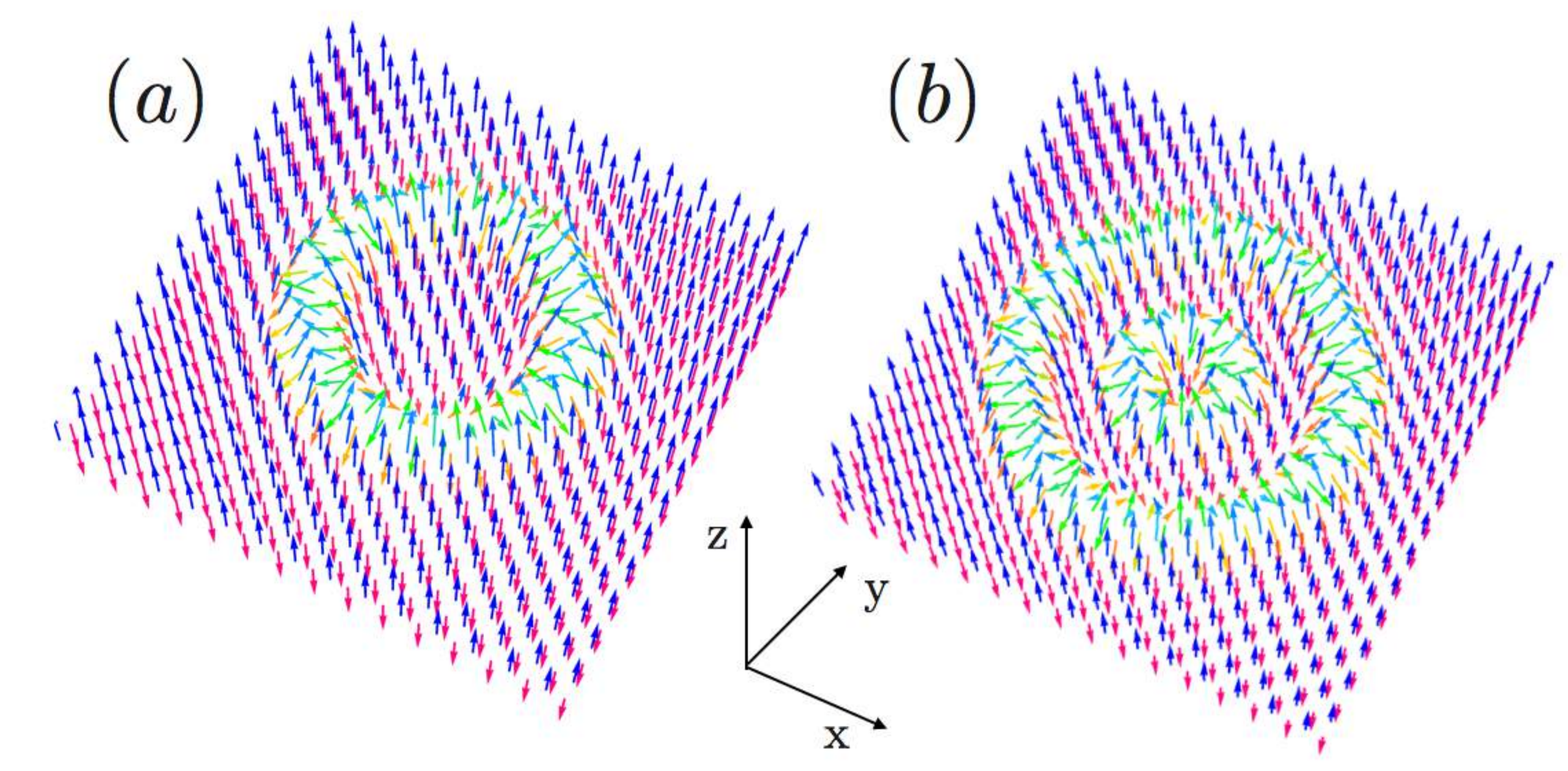}
       \caption{Schematics of a N\'{e}el-type antiferromagnetic (a) skyrmion and (b) skyrmion duplex. There are two magnetic sublattices in N\'{e}el ordered states in the square lattice, and a skyrmion and skyrmion duplex can be seen as bound states of their ferromagnetic counterparts living in different magnetic sublattices. Just as Fig.~\ref{schematics}, the colors of arrows represent the $z$ component of spins. }
          \label{AFMSk_and_SkD}
    \end{figure}

We use the following canonical model of two-dimensional chiral AFMs on a square lattice
\begin{align}
H_{AF} &= J \sum_{\vec{r}} \vec{m}_{\vec{r}}\cdot\left(\vec{m}_{\vec{r} + a \vec{e}_{x}}+ \vec{m}_{\vec{r} + a \vec{e}_{y}} \right) 
+\sum_{\vec{r}}\vec{D}_{i}\cdot\left(	\vec{m}_{\vec{r}}\times \vec{m}_{\vec{r}+a\vec{e}_{i}}		 \right)
-A\sum_{\vec{r}}(m^z_{\vec{r}})^2,
\label{model_AFM}
\end{align}
where a spin localized at $\vec{r}$ is represented as $\vec{m}_{\vec{r}}$ with its norm normalized to unity.
The exchange coupling is antiferromagnetic ($J > 0$) and $A$ is the anisotropy energy along the $z$ axis. The DM interaction stabilizes antiferromagnetic skrmions in this model. For example, if we choose $\vec{D}_{x} = D \vec{e}_{y}$, $\vec{D}_{y} = -D \vec{e}_{x}$, a N\'{e}el-type antiferromagnetic skyrmion [shown in Fig.~\ref{AFMSk_and_SkD}(a)] appears. The study of such antiferromagnetic solitons in noncentrosymmetric two-dimensional AFMs itself has a rather long history~\cite{BaryakhtarJETP,Wolf2002}, but recently, their presence in the field of magnetism is rapidly increasing in the context of antiferromagnetic spintronics~\cite{Zhang:2016aa,PhysRevLett.116.147203,1367-2630-18-7-075016}.

\begin{figure}[htbp]
   \centering
\includegraphics[width = 120mm]{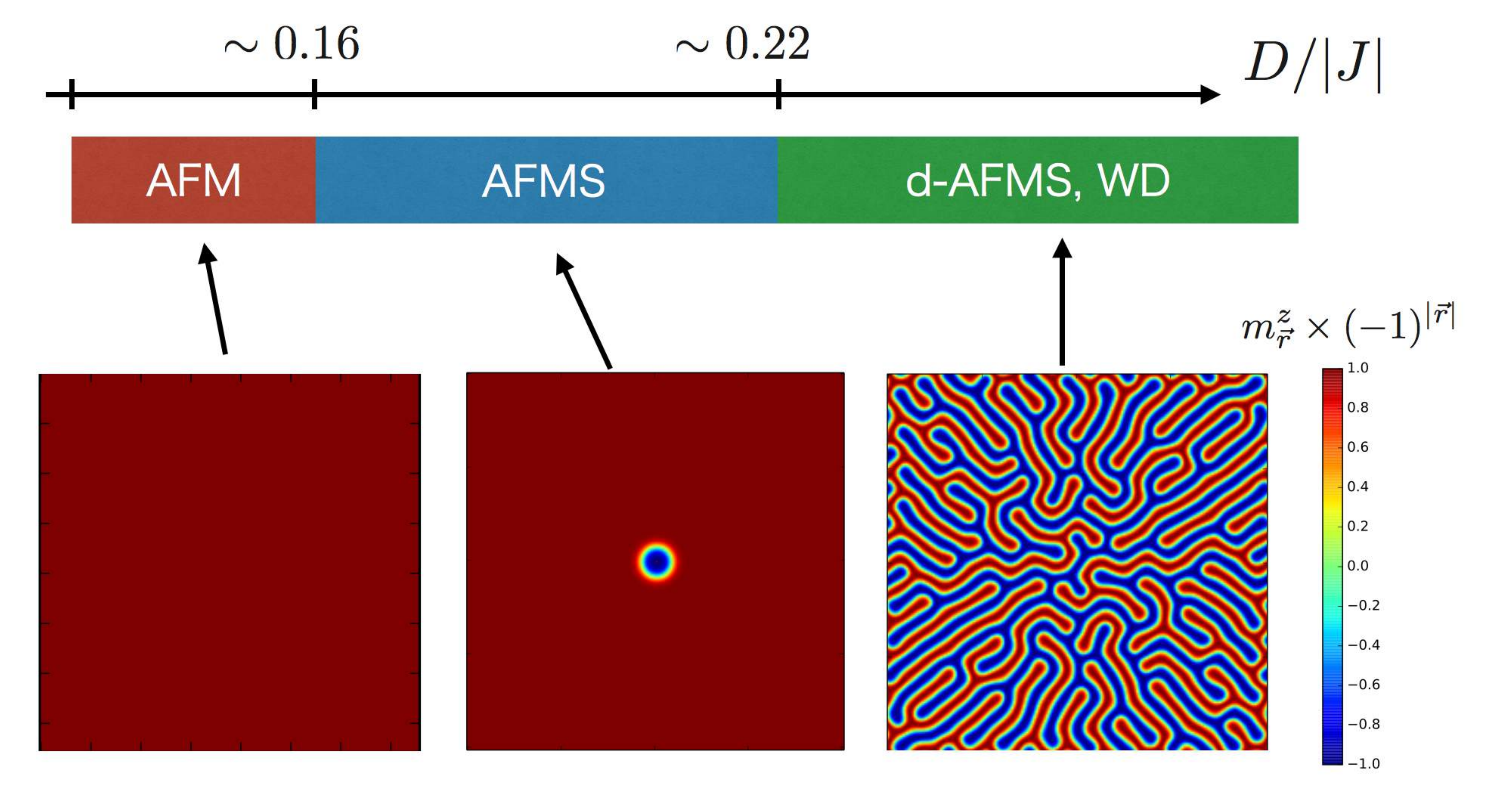}
       \caption{Phase diagram of the canonical model of chiral AFMs Eq.~\eqref{model_AFM} for $A/J = 0.055$. When DM interaction is very weak, we have an antiferromagnetic (AFM) region where we cannot have skyrmions. As we increase DM interaction $D$, antiferromagnetic skyrmions become energetically stable at $D/J \sim 0.16$ as isolated defects. For larger DM interaction ($D/J \ge 0.22$), skyrmions deform to lower their energy (d-AFM state) and eventually warm domains are formed (WD). The phase boundary between d-AFM and WD is unclear from our calculations. We visualize the spin texture of typical states in each phase obtained from LLG calculations by using staggered spins $m^z_{\vec{r}}\times(-1)^{|\vec{r}|} \equiv m^z_{\vec{r}=(i,j)}\times(-1)^{i+j}$.}
          \label{AFM_phasediagram_gs}
    \end{figure}

When $A/J = 0.055$, the phase diagram of this model~\cite{Zhang:2016aa} with $\vec{D}_{x} = D \vec{e}_{y}$, $\vec{D}_{y} = -D \vec{e}_{x}$ is given in Fig.~\ref{AFM_phasediagram_gs}. For $D/J < 0.22$, the ground state is a N\'{e}el state, but we can divide this phase into two regions. In the region with $D/J < 0.16$, antiferromagnetic skyrmions are unstable while if $D/J>0.16$, they are energetically stable as isolated topological defects with a very long lifetime. Following Ref.~\cite{Zhang:2016aa}, we call the latter region an antiferromagnetic skyrmion (AFMS) region. For larger DM interaction, skyrmions deform (d-AFMS region) and eventually turn into warm domains (WD region). In lower panels of Fig.~\ref{AFM_phasediagram_gs} we show typical states in these phases as staggered spins $m^z_{\vec{r}}\times(-1)^{|\vec{r}|} \equiv m^z_{\vec{r}=(i,j)}\times(-1)^{i+j}$. We note that WDs can be regarded as a collection of strongly deformed skyrmions, so that the boundary between the d-AFMS region and the WD region is unclear from our LLG calculations in a finite size system. 


\subsection{Antiferromagnetic skyrmion multiplex}
We focus on the AFMS region in Fig.~\ref{AFM_phasediagram_gs} and take $\vec{D}_{x} = D \vec{e}_{y}$, $\vec{D}_{y} = -D \vec{e}_{x}$ in the following. We treat the effect of applied vortex beams by the sLLG equation, assuming temperature proportional to the beam intensity. Then we observe N\'{e}el-type antiferromagnetic SkDs, whose spin texture is defined in Fig.~\ref{AFMSk_and_SkD}(b), after the irradiation in the proper manner (see Sec.~\ref{sec: AFM_opt}). Observing the spin texture carefully, we find that an antiferromagnetic SkD consists of two ferromagnetic SkDs in the two magnetic sublattices. Antiferromagnetic SkDs are energetically stable and their lifetime is longer than our calculation periods. Just as in the ferromagnetic case, there can appear a family of stable ring-shaped defects, antiferromagnetic skyrmion multiplexes (corresponding objects of ferromagnetic ones).  Besides, as we will see below, in chiral AFMs, even the simplest vortex beam with $p = 0$ can create a variety of topological defects other than skyrmionium.
In the case of AFMs, experimental observation of their spin textures with Lorentz TEM is difficult. In Ref.~\cite{PhysRevLett.116.147203}, instead, neutron scattering and x-ray magnetic linear dichroism are proposed as a probe of antiferromagnetic skyrmions. We expect that these methods are also applicable to antiferromagnetic skyrmion multiplexes.


   \begin{figure}[htbp]
   \centering
\includegraphics[width = 150mm]{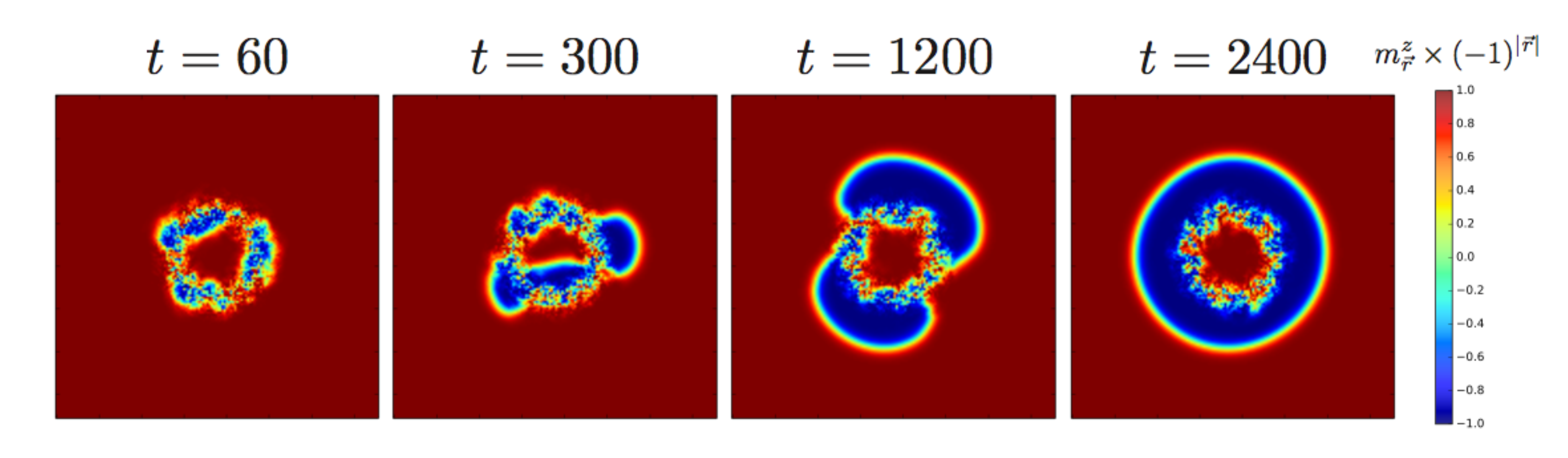}
       \caption{Time evolution of staggered spins $m^z_{\vec{r}}\times(-1)^{|\vec{r}|} \equiv m^z_{\vec{r}=(i,j)}\times(-1)^{i+j}$ for a particular trial with parameters $D/J = 0.205, A/J=0.055, T_{0}/J = 1$, $w = 12.5 a$, and $\alpha = 0.1$. The ring-shaped heating caused by vortex beams with $p =0$ and $m = 5$ creates magnetic domains different from the background N\'{e}el order. These domains merge to form an antiferromagnetic skyrmion duplex. A time step corresponds to 0.7 ps for $J = 1$ meV.}
          \label{AFM_SkD_formation}
    \end{figure}
   

   \begin{figure}[htbp]
   \centering
\includegraphics[width = 150mm]{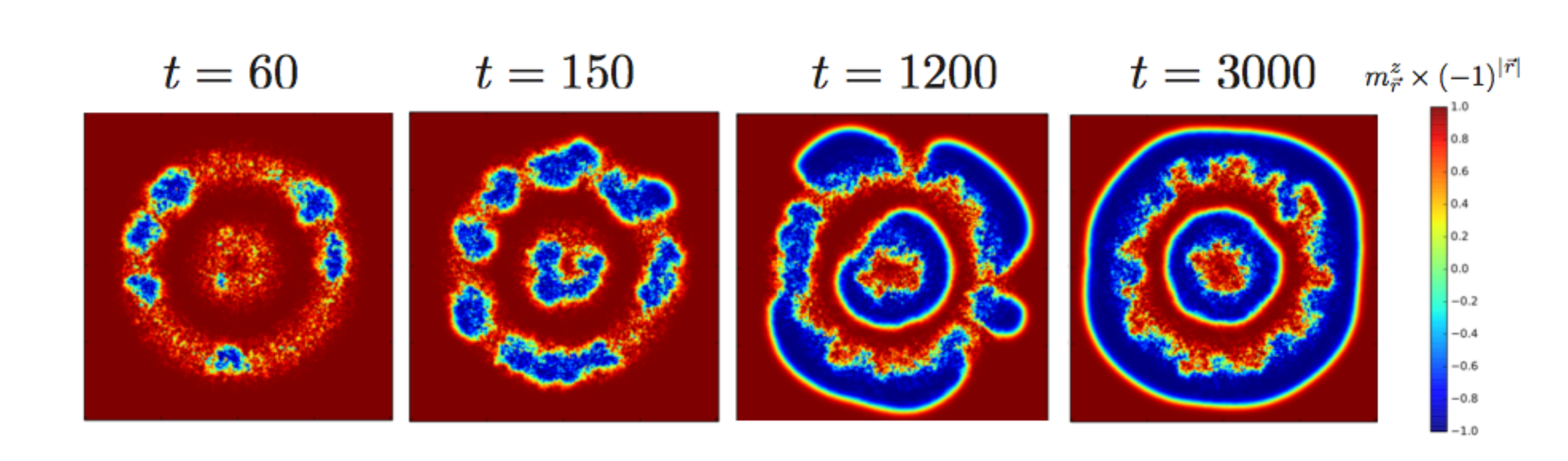}
       \caption{Time evolution of staggered spins $m^z_{\vec{r}}\times(-1)^{|\vec{r}|} \equiv m^z_{\vec{r}=(i,j)}\times(-1)^{i+j}$ for a particular trial for $D/J = 0.21, A/J=0.055$, and $\alpha = 0.1$. We superimpose two vortex beams ($m = 1$ and $m = 20$) with $T_{0}/J = 0.5$ and $w = 20 a$, to realize temperature with a double-ring profile. We see that the heating leads to formation of an antiferromagnetic skyrmion quadplex. The system is a periodic square lattice with 200 $\times$ 200 sites.}
          \label{AFM_triplex_formation}
    \end{figure}
    

\subsection{Formation of skyrmion multiplex}
 Following the analysis in the ferromagnetic case, we study the time evolution of spins under the beam-induced temperatures. The sLLG equation for the antiferromagnetic Hamiltonian~\eqref{model_AFM} is
 \begin{align}
\frac{d\vec{M}_{\vec{r}}}{dt} &= - \gamma \vec{M}_{\vec{r}} \times \left(-\frac{\partial H_{AF}}{\partial \vec{M}_{\vec{r}}} +  \vec{h}_{T(\vec{r})}(t)\right)+ \alpha\frac{\vec{M}_{\vec{r}}}{|\vec{M}_{\vec{r}}|}\times \frac{d\vec{M}_{\vec{r}}}{dt},
\label{sLLG-AFM}
\end{align}
where $\vec{M}_{\vec{r}} = \hbar \gamma \vec{m}_{\vec{r}}$, and $\vec{h}_{T(\vec{r})}(t)$ is again the random field satisfying Eq.~\eqref{random} and $\sigma(\vec{r}) = 2 k_{B}T(\vec{r})\alpha/(\gamma^2\hbar)$. We take a system with $150\times150$ sites and impose periodic boundary condition. The time step of the Heun method is set to $\Delta t = 0.03$. Hereafter we take $J = 1$ and fix $D = 0.205$ and $A= 0.055$. After several trials and errors, we find that it is advantageous to keep the temperature constant for long periods to achieve a high success probability of defect creation. Therefore we assume the time dependence of the temperature induced by the vortex beams as $T(t) = T_0 \Theta(t_{0}-t) \Theta(t)$ [shown in Fig.~\ref{AFM_phasediag}(d)] and examine the probability for several values of $t_{0} = 3000, 5000$, and $7000$. Here $T_{0}$ is set to be proportional to the local intensity of vortex beams with $p = 0$ and $m = 5$, and $\Theta(x)$ is the step function. In Fig.~\ref{AFM_SkD_formation}, we show the time evolution within a particular trial. We see that under the static heating, domains of a N\'{e}el state different from the background one grow. These domains merge and result in an antiferromagnetic SkD. We also show the formation process of an antiferromagnetic skyrmion quadplex in Fig.~\ref{AFM_triplex_formation}. Here, instead of using vortex beams with $p = 1$, we superimpose two vortex beams ($p  = 0$) with different ring size to achieve the double-ring-shaped temperature profile.

\subsection{Optimal way of applying vortex beams}\label{sec: AFM_opt}
Here we optimize the way we apply vortex beams to achieve high success probability of creating antiferromagnetic SkDs. We try calculations 20 times for each set of $(T_{0}, t_{0}, w)$ and obtain the success probability in the same way as the ferromagnetic case for vortex beams with $m = 5$ and $p = 0$. The numerical calculations are performed in the AFMS region of chiral AFMs where the ground state is a N\'{e}el ordered state. Therefore, we are choosing a stable N\'{e}el state as the initial state of our calculations. This is in contrast with the ferromagnetic case where we assumed the metastable ferromagnetic initial state. 

\begin{figure}[htbp]
   \centering
\includegraphics[width = 150mm]{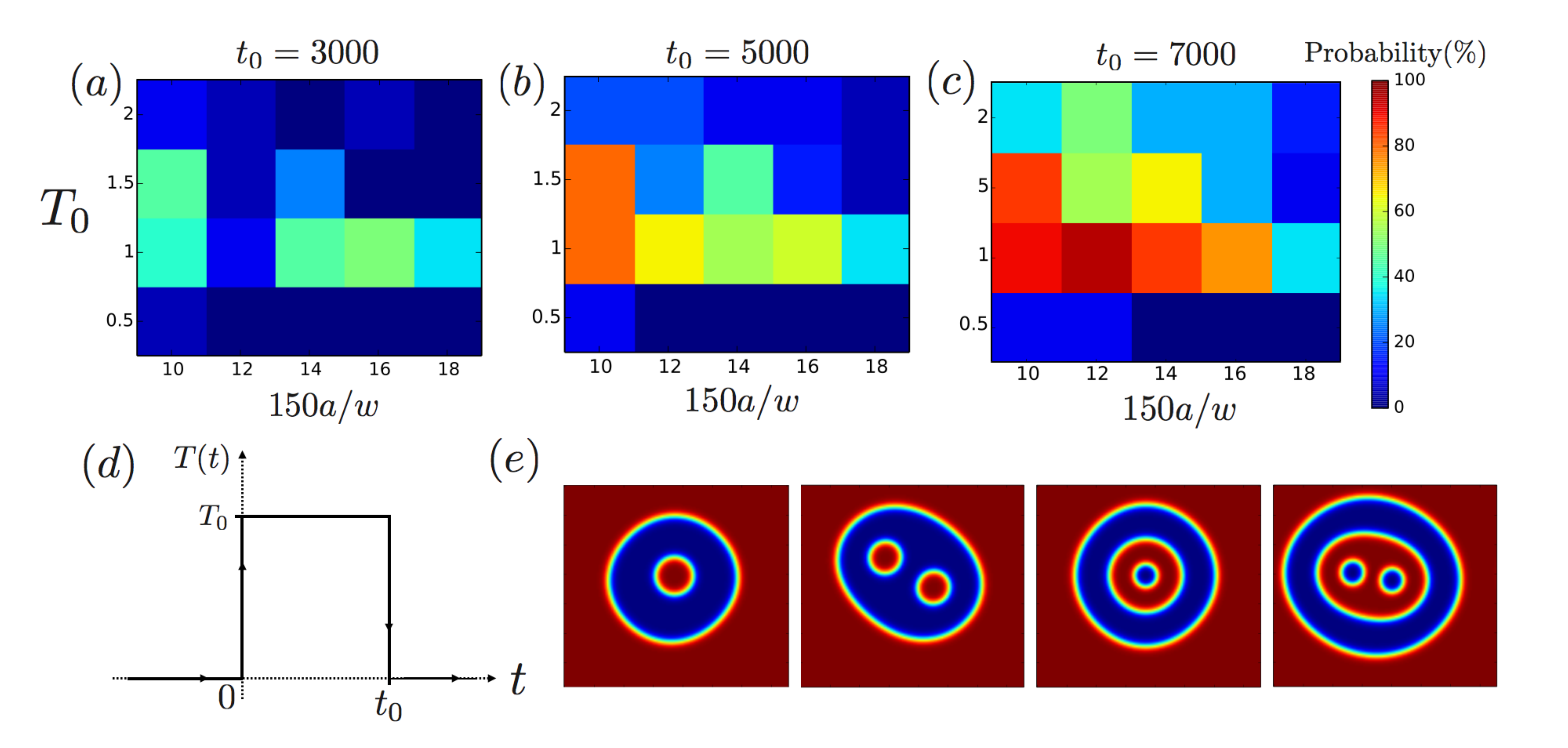}
       \caption{Success probability of creating antiferromagnetic skyrmion multiplexes by vortex beams with $p = 0$ and $m = 5$. We fix $J = 1$, $D = 0.205$, $A = 0.055$, and $\alpha = 0.1$, with which the system is in the AFMS region (see Fig.~\ref{AFM_phasediagram_gs}). The initial state at $t = 0$ is a N\'{e}el ordered ground state and the temperature is varied in accord with (d) for (a) $t_{0} = 3000$, (b) $t_{0} = 5000$, and (c) $t_{0} = 7000$. The probability is high when the wavelength of the vortex beams is of the same order of the size of antiferromagnetic skyrmions, and the period of the irradiation of the vortex beams is long. (e) Skyrmioin multiplexes observed after the irradiation of vortex beams. We show the $z$ component of staggered spins for visibility.}
          \label{AFM_phasediag}
    \end{figure}

Contrary to the ferromagnetic case, here even the simplest vortex beam with $p = 0$ can create various  topological defects other than skyrmions and SkDs though not well controlled. In Fig.~\ref{AFM_phasediag}(e) we show the $z$ component of staggered spins for several examples of topological defects we observed, including that of an SkD. In the case of chiral FMs, due to the external magnetic field $B_{z}$, complicated magnetic structures are energetically unfavored and only SkDs are observed within our calculations for the vortex beam with $p = 0$. However, since there is no corresponding ``staggered magnetic field" in the present model, two possible N\'{e}el states are energetically degenerate. Therefore, even if the spin texture of a topological defect is intricate, it costs energy only at the domain boundaries and can be energetically stabler than that in chiral FMs (the same situation as Ref.~\cite{PhysRevLett.110.177205}). All these defects have ring-shaped structures and reflect the spatial profile of our vortex beams. We confirm that they have a lifetime longer than our calculation periods and call them also a skyrmion multiplex.

The obtained success probability of creating defects for vortex beams with $p=0$ and $m = 5$ is summarized in Fig.~\ref{AFM_phasediag}a. Here we regard each trial to be in success if one of ring-shaped defects [some examples are shown in Fig.~\ref{AFM_phasediag}(e)] is generated. The optimal beam waist is found to satisfy $w \sim 11.5 a$. If the lattice constant is $a = 5 \mathrm{\AA}$ and the beam is focused well $w \sim 0.5 \lambda$, the wavelength $\lambda \sim 12$ nm, again close to the size of (antiferromagnetic) skyrmions and in the EUV region. Just as in the ferromagnetic case, we expect that the wavelength can be lifted by using materials with larger antiferromagnetic skyrmions.\par


\section{Conclusion}\label{sec: concl}
We proposed an application of optical vortex~\cite{PhysRevA.45.8185} or electron vortex beam~\cite{PhysRevLett.99.190404} to magnetism. These vortex beams carry orbital angular momentum and have a ring-shaped profile of intensity. We numerically demonstrated that the spatial pattern of those beams can be ``printed" to chiral magnets as a class of ring-shaped topological magnetic defects.

Considering the mismatch between time and length scales of chiral magnets, we modeled the effect of vortex beams as heating. The heating realizes spatially modulated temperature proportional to the local intensity of the  beams.  By solving the stochastic Landau-Lifshitz-Gilbert equation~\cite{PhysRevB.58.14937} for canonical models of chiral ferro-~\cite{PhysRevLett.108.017601} and antiferromagnets~\cite{Zhang:2016aa}[Eqs.~\eqref{model} and \eqref{model_AFM}], we found that the ring-shaped profile of the vortex beams, which originates from their nonvanishing orbital angular momentum, can be transferred to chiral magnets as topological defects. We confirmed that by changing the beam parameters we can create a variety of topological defects including ordinary skyrmions~\cite{Fert:2013aa} and general $n\pi$ vortices (skyrmion multiplexes). We confirmed that skyrmion multiplexes in chiral FMs are energetically stable in the skyrmion crystal phase. That is, they are stable within the framework of our LLG calculations. For chiral AFMs, they are stable in the broad region of the phase diagram reflecting the degeneracy of N\'{e}el states.

We gave optimized way of applying vortex beams to achieve high probability of creating skyrmion multiplexes in chiral magnets (Fig.~\ref{phasediag_alpha=0.1} for the ferromagnetic case and Fig.~\ref{AFM_phasediag} for the antiferromagnetic case). The appropriate wavelength of vortex beams is found to be of the same order of the size of skyrmions. Typically the size of skyrmions in chiral magnets is O$(1)$$\sim$O$(10)$ nm so that the proper wavelength is in the extreme ultraviolet region. However, recently much larger skyrmions~\cite{Jiang283,Yu_2016,largesk,Woo:2016aa,Moreau-LuchaireC.:2016aa,Boulle:2016aa} have been observed in chiral magnetic films in which the optimal wavelength would be lifted to the visible light region. Moreover, we can utilize recent developments in plasmonics~\cite{Heeres:2014aa,PhysRevX.2.031010, tanaka} to overcome this mismatch issue.

Our method using vortex beams has the advantage that the approach is applicable to both metallic and insulating chiral magnets and works for both chiral ferro- and antiferro- magnets in the identical manner. Moreover, by changing beam parameters it is also possible to create a general $n\pi$ vortex. 
These properties stand in strong contrast with the strategies using current pulses~\cite{PhysRevB.85.174416,Romming636,PhysRevB.94.094420}, where their applicability is so far limited to metallic chiral ferromagnets and to the generation of a skyrmion or a 2$\pi$ vortex.

\section{Acknowledgement}
We thank Hideki Hirori, Koichiro Tanaka, Yasunori Toda, Hidekazu Misawa, and Yasuhiro Tada for useful comments. HF is supported by Advanced Leading Graduate Course for Photon Science (ALPS) of Japan Society for the Promotion of Science (JSPS) and JSPS KAKENHI Grant-in-Aid for JSPS Fellows Grant No.~JP16J04752. MS is supported by JSPS KAKENHI Grant-in-Aid for Scientific Research(A) No.~JP15H02117.
This research was supported in part by the National Science Foundation under Grant No. PHY11-25915. 
\bibliography{vortex}
\end{document}